\documentclass[preprint,aps,showpacs,showkeys]{revtex4}
\usepackage{graphicx,amssymb}

\begin{document}
\title{Magnetic Behavior of Single La$_{0.67}$Ca$_{0.33}$MnO$_3$ Nanotubes: Surface and Shape Effects}

\author{M. I. Dolz}
\affiliation{Centro At\'omico Bariloche, Comisi\'on Nacional de
Energ\'{\i}a At\'omica, Av. Bustillo 9500, R8402AGP S. C. de
Bariloche, Argentina}

\author{W. Bast}
\affiliation{Centro At\'omico Bariloche, Comisi\'on Nacional de
Energ\'{\i}a At\'omica, Av. Bustillo 9500, R8402AGP S. C. de
Bariloche, Argentina}

\author{D. Antonio}
\affiliation{Centro At\'omico Bariloche, Comisi\'on Nacional de
Energ\'{\i}a At\'omica, Av. Bustillo 9500, R8402AGP S. C. de
Bariloche, Argentina}

\author{H. Pastoriza}
\affiliation{Centro At\'omico Bariloche, Comisi\'on Nacional de
Energ\'{\i}a At\'omica, Av. Bustillo 9500, R8402AGP S. C. de
Bariloche, Argentina}

\author{J. Curiale}
\affiliation{Centro At\'omico Bariloche, Comisi\'on Nacional de
Energ\'{\i}a At\'omica, Av. Bustillo 9500, R8402AGP S. C. de
Bariloche, Argentina}

\author{R. D. S\'anchez}
\affiliation{Centro At\'omico Bariloche, Comisi\'on Nacional de
Energ\'{\i}a At\'omica, Av. Bustillo 9500, R8402AGP S. C. de
Bariloche, Argentina}

\author{A. G. Leyva}
\affiliation{Centro At\'omico Constituyentes, Comisi\'on Nacional
de Energ\'{\i}a At\'omica, Av. Gral Paz 1499 (1650) San
Mart\'{\i}n, Argentina}

\date{\today}

\begin{abstract}

We report magnetization experiments in two magnetically isolated
ferromagnetic nanotubes of perovskite
La$_{0.67}$Ca$_{0.33}$MnO$_3$. The results show that the magnetic
anisotropy is determined by the sample shape although the coercive
field is reduced by incoherent magnetization reversal modes. The
temperature dependence of the magnetization reveals that the
magnetic behavior is dominated by grain surface properties. These
measurements were acquired using a Silicon micro-mechanical
oscillator working in its resonant mode. The sensitivity was
enough to measure the magnetic properties of these two samples
with a mass lower than 14 picograms and to obtain for the first
time the magnetization loop for one isolated nanotube.

\end{abstract}


 \maketitle

\section{INTRODUCTION}

Manganites are complex oxides that adopt a pseudo-cubic perovskite
crystal structure. The discovery of colossal magnetoresistance in
films of these materials \cite{Jin94} was a great leap for the
field of spintronics. The first impact was its use as electrodes
in magnetic tunnel junctions \cite{Lu96}, giving tunnel
magnetoresistance ratios one order of magnitude larger than of
those obtained with transition-metal electrodes. The fabrication
of manganite nanotubes are nowadays on the rise due to their
several possible technological applications. These nanotubes could
impact on technology in different ways. One possible application
is in solid-oxide fuel cells. Manganites nanotubes make good
cathodes thanks to their granular structure where gases may be
efficiently distributed and because they conduct both electrons
and oxygen ions, and are resistant to high-temperature oxidizing
environments \cite{Leyva06}. Another very important possible
application of metallic manganites nanotubes in the field of
nanotechnology is as highly localized sources of electrons
possessing spins of a particular orientation, which can be
perfectly aligned \cite{Bowen03}. Manganite also exhibit
challenging electronic and magnetic properties because of their
tendency to present coexistence of different phases in a wide
range of scale lengths \cite{Mathur03}. The relative quantity of
phases coexisting can be modified by means of small external
forces due to the strong competition between the present
interactions. The ability to fabricate nanotubes (NT) of these
compounds has recently been demonstrated \cite{levy03}. These
nanostructures are usually studied by measurements in large sets
where the intrinsic response of individual structures is clouded
by the global response. Given the necessity of having a good
characterization and knowledge of each of these complex
nanostructures this work is oriented towards the study of
individual La$_{0.67}$Ca$_{0.33}$MnO$_3$ (LCMO) NT magnetic
properties.

In this paper, we show that the saturation magnetization
temperature dependence obtained for two isolated NT is clearly
different from the bulk sample. We observe a lineal temperature
dependence. This behavior is explained having in mind the
predominating surface effects of each grain that constitutes a NT
\cite{sanchez96,kaneyoshi,park98}. This granular structure favors
the existence of weak collective modes for the magnetization
reversal that depress NT coercive fields ($H_{c}$) respect to that
expected by the shape anisotropy constant. But, the complete
magnetization loop obtained for one single NT shows a little
increment of $H_{c}$ respect to those measured in a powder of NT
and in a bulk sample \cite{curiale07}. One of the main
achievements of our work has been to obtain for the first time the
complete loop of magnetization of one isolated NT.

On the subject of measuring magnetic properties of small samples,
several approaches have been reported in the literature:
microSQUIDs \cite{wernsdorfer96}, which are very sensitive but
constrained to low fields and low temperatures; Alternate Gradient
Magnetometers (AGM) \cite{zijlstra70} have also demonstrated high
sensitivity \cite{todorovic98} but quantitative measurements are
difficult to obtain since the measured signal strongly depends on
the sample shape and on the exact location of the sample in the
field gradient \cite{barbic04}. Central to our approach is the use
of Micro-mechanical torsional oscillators as magnetometers. These
systems fabricated with MEMS technology, extend all the advantages
of high-$Q$ mechanical oscillators measurements \cite{kleiman85}
to the microscopic scale and, before focusing on the results of
this paper, we will explain how it is possible to measure magnetic
properties with these MEMS.

\section{EXPERIMENTAL}

In our experiments we used a variant of the micro-torsional
oscillators presented by Bolle {\it et al.} \cite{bolle99}. These
poly-silicon oscillators were fabricated in the MEMSCAP
\cite{memscap} foundry using its Multiuser process (MUMPS). Each
oscillator consists of a 44 $\times$ 106 $\mu$m$^{2}$ released
plate anchored to the substrate by two serpentine springs. Fixed
to the substrate, underneath the plate,  two electrodes are used
for the driving and motion detection. A 100 mV peak-to-peak ac
voltage is applied between one electrode and the plate to induce
the motion, and the change in capacitance is detected in the other
electrode using a Phase Sensitive Detection scheme tuned at twice
the driving frequency. A more detailed description of the
experimental detection setup can be found elsewhere \cite{dolz07}.

A powder of NT of LCMO was fabricated by the pore-filling method
\cite{leyva04}. With this procedure NT are build from grains of 25
nm in diameter \cite{curiale07}. A small amount of the obtained
conglomerated NT was placed on a lithographically patterned
substrate where individual NT were identified and selected using a
scanning electron microscope (SEM). The chosen NT were placed on
top of a silicon micro-oscillator \cite{dolz07} using hydraulic
micro-manipulators under an optical microscope, and were glued to
it with a sub-micrometer drop of Apiezon$^\copyright$ N grease. In
Fig. 1(b) we show a scanning electron micrograph of the two NT
glued on top of the MEMS oscillator. All the results reported in
this paper correspond to measurements done in two NT of 700 nm in
diameter and 9.5 $\mu$m in length placed perpendicular to the
rotation axis of the oscillator and separated 40 $\mu$m (See Fig.
1(b)). At this distance the interaction between the NT is
negligible as the dipolar magnetic field, estimated from the
expected maximum magnetization value, is slightly larger than the
earth magnetic field and two orders of magnitude lower than the
necessary field to reverse the magnetization of the NT. The
measurements were taken in vacuum inside a closed-cycle
cryogenerator where the temperature can be varied between $14$ and
$300\,$K. The magnetic field was provided by a split electromagnet
that can be rotated in the plane perpendicular to the axis of
rotation of the oscillators with an accuracy of 1$\,^{\circ}$.

The experiment consists in measuring the oscillator`s torsional
mode resonant frequency as a function of magnetic field and
temperature. This is accomplished by sweeping the driving
frequency and detecting the oscillator amplitude. The measured
amplitude is squared and fitted with a lorentzian function, from
which the resonant frequency and quality factor are obtained. The
natural resonant frequency ($\nu_0$) of an oscillator in the
torsional mode is given by: $ 2\pi\nu_0=\sqrt{\frac{k_e}{I}},$
where $k_e\simeq 7.82\times10^{-3}\,\mathrm{dyn}\cdot\mathrm{cm}$
is the elastic restorative constant of the serpentine springs and
$I=3.8\times10^{-14}\,\mathrm{g}\cdot\mathrm{cm}^2$ is the plate's
moment of inertia. In our oscillators this mode has a resonant
frequency close to 72200 Hz and a quality factor $Q$ greater than
$5\times 10^{4}$, which means that the width of the resonant peak
is less than 2 Hz.

When we attach a magnetic sample to the oscillator the resonant
frequency $\nu_r$  changes to:
\begin{equation}
2\pi \nu_{r} = \sqrt{\frac{k_e+k_M}{I}},
\end{equation}
where $k_M$ is the variation in the effective elastic constant
originated by the magnetic interaction between the sample and the
external magnetic field. For the experimental conditions $I$ does
not change during the measurements and $\Delta\nu << \nu_0$.
Therefore it is possible to express
\begin{equation}
k_M \simeq 8 \pi^2 I \nu_0 \Delta\nu,
\end{equation}
where $\Delta\nu$ is the change in the resonant frequency. This
change as a function of a magnetic field applied parallel to the
principal axis of the NT at different temperatures is showed in
Fig.~2(a). These results show that the oscillator is sensitive
enough to detect the NT  magnetic response.

In our experiment the oscillatory motion produces a tilt between
the fixed magnetic field and the NT axis. At high magnetic fields,
when the magnetization of the sample is saturated, $M=M_s$ and
$\frac{dM}{dH}=0$, the change in the resonant frequency can be
written as \cite{morillo98}:
\begin{equation}
\frac{1}{8\pi^2I\nu_0\Delta\nu}\simeq
\frac{1}{k_M}=\frac{1}{2KV}+\frac{1}{M_S V H_0}
\end{equation}
where $\Delta\nu$ is the change in the resonant frequency at a
given magnetic field, $V$ is the sample volume, and
$K=\frac{1}{2}N M_S^2$ is the shape anisotropy energy density.
From a linear square fit of the high magnetic fields data the
values for $M_S V$ and $K$ are obtained. In Fig.~2(b) we plot
$\Delta\nu^{-1}$ vs $H_0^{-1}$ for the data taken at some selected
temperatures showing the excellent correlation obtained with this
linear fit in a wide range of fields. $K$ depends on temperature
and its extrapolated value at $T=0$K is $1.24\times10^{-6}$ erg.

\section{RESULTS AND DISCUSSION}

The saturation magnetization temperature dependence of our NT
obtained with the described procedure is plotted in Fig.~3. For
comparison we have plotted in the same graph the results obtained
for a $0.40$ mg piece of sintered LCMO taken in a commercial SQUID
magnetometer with an applied field of $10\,$kOe. Clearly the
temperature dependence of the individual NT magnetization differs
from the bulk ferromagnet.

Magnetization measurements done in samples of varying sizes
\cite{sanchez96} have shown that the diminution of grain size is
associated with a decrease on the magnetization and a change in
the temperature dependence. The sample size reduction implies an
increase on the surface to volume ratio which means that surface
effects become more relevant to describe the physical behavior of
the sample. At the surface of crystalline grains the atomic
coordination is reduced and atomic disorder is much more important
than in the bulk. In Manganese-based perovskites the magnetic
properties result from the interplay of many complex phenomena.
The ferromagnetic double-exchange spin--spin coupling competes
with an anti-ferromagnetic super exchange. Both are very sensitive
to the Mn-O-Mn bond angle and distance. This implies that the
magnetization at the surface of these compounds is highly
susceptible to surface conditions. The magnetic modifications at
the surface are usually argued as the origin of a magnetic dead
layer, but in general it affects in a more complex way the
magnetic properties. The detailed description of the surface
magnetism temperature dependence is strongly dependent on the
surface local conditions. In a mean field calculation it has been
shown \cite{kaneyoshi} that the surface magnetization has a linear
temperature dependence. Photo-emission measurements of the surface
magnetization performed in other manganese perovskite compounds
\cite{park98} show this linear temperature dependence, as obtained
in our experiments. Taking into account that our NT are built-up
from grains of 25 nm in diameter \cite{curiale07} and the surface
magnetism could extend 2 nm in depth,  approximately 50 \% of the
magnetic moments are weakly correlated on the surface and dominate
the global magnetic behavior.

The volume for the two NT is $2.32\times10^{-12}\,$cm$^3$, it was
estimated from the external dimensions measured in a SEM and
assuming a nominal wall thickness of $60\,$nm \cite{curiale07}.
Considering this volume and the bulk manganite density
$6.03\,$g/cm$^3$ \cite{sanchez98}, their total mass is about
$14\times10^{-12}\,$g. In consequence, the saturation
magnetization per mass extrapolated to $0\,$K is 52 emu/g (See
Fig.~3) which is greater than the value obtained for a powder of
these NT \cite{curiale07} and is less than the bulk LCMO value
($98\,$emu/g). The difference with the bulk`s value can be assumed
by the existence of a {\em magnetic dead layer} in each grain that
constitutes the wall of the NT. Considering that the ratio between
the obtained saturation magnetization and the bulk value ($0.53$)
should be equal to the ratio between the magnetic core and the
total grain volume (with an average diameter of $26\,$nm
\cite{curiale07}), we can estimate the width of this magnetic dead
layer. Its estimated value is around $2\,$nm and is very close to
the $1.6\pm0.4\,$nm value obtained from the saturation
magnetization of a powder of randomly oriented NT. Due to the
granular morphology of the NT, the dead layer thickness obtained
from our results must be taken as a superior limit. The NT density
is smaller than the bulk, which means that the real mass of the NT
is lower than calculated and the dead layer is smaller.

In order to obtain the full magnetization loop in our experiments
it must be noted that the restorative constant $k_M$ generated by
the magnetic sample has to be evaluated through the second
derivative of the magnetic free energy respect to the displacement
angle $\theta$ between the sample and the magnetic field. The
magnetic energy density for a ferromagnetic sample with uniaxial
anisotropy in a magnetic field can be described by:
\begin{eqnarray}
E=&&-{\bf\vec{M}\cdot\vec{H}_0}+\frac{1}{2} N |\bf{\vec{M}}|^2 \sin^2(\varphi)  \nonumber \\
=&&-MH_0\cos(\theta-\varphi)+\frac{N M^2}{2}\sin^2(\varphi)
\end{eqnarray}
where $\vec{M}$ is the sample magnetization, $\vec{H_0}$ is the
external magnetic field, $N$ is the difference between the
demagnetization factors of the  magnetic hard and easy directions,
and $\varphi$ is the angle between the magnetization vector and
the easy direction of the sample (See Fig.~1(a)). This angle has
to be calculated from the evaluation of
$\frac{dE}{d\varphi}|_\theta=0$  for a given $\theta$. Taking into
account that $M$  depends on $H=H_0-H_D$ ($H_D$ the demagnetizing
field) and $\varphi$ depends on $\theta$, for $\theta$ close to
zero:
\begin{eqnarray}
\frac{1}{V} k_M=&&\frac{d^2E}{d\theta^2} \\ \nonumber = && \frac{
MNH_0\left[M(MN+H_0)+\frac{dM}{dH}H_0(MN+2H_0) \right] }{
(MN+H_0)^2}.
\end{eqnarray}
The complete magnetization loop can be obtained from the measured
$k_M$ data through  a nonlinear least-squared fit. In this
procedure we used $M$ and $\chi=\frac{dM}{dH}$ as free field
dependent parameters and $N$ as a fixed parameter obtained from
the fitting of the data at high fields.

In Fig.~4 we plot $m(H)$ obtained from our measurements at 14 K.
In the same graph we have plotted the hysteresis loop for a $1.84$
mg sample of NT powder measured in a SQUID magnetometer at the
same temperature. There is an excellent agreement between both
measurements. In the randomly oriented NT powder the measurement
shows an S shaped curve originated from the distribution of
anisotropies in the sample. In contrast the isolated NT
measurement is more abrupt, as expected.

More information can be extracted from the raw $\Delta\nu$ vs $H$
data. From equation~(5) we have $\Delta\nu(H_0)=0$ when $H_0=0$ or
$M(H_0)=0$, the later corresponding to the $H_{c}$. These
zero-crossing points are visible in the data presented in the
inset of Fig.~2(a). The values obtained for the $H_{c}$ ($\approx
350\,$Oe) are much smaller than those expected for a $H_{c}$
produced by the shape anisotropy constant obtained from our data
($\approx 3320\,$Oe), but are slightly larger than those measured
in a powder of NT (as shown in Fig.~4). Considering the granular
structure of our NT and a weak magnetic interaction between
grains, the existence of weak collective modes for the
magnetization reversal could be favored (i.e. as fanning or
buckling of the magnetic moments) which would result in a
depressed NT $H_{c}$.

\section{CONCLUSIONS}

In conclusion, we have presented magnetization measurements of
single LCMO granular NT using a silicon micro-oscillator. Thanks
to the oscillator's high Q factor and soft restorative constant we
have obtained a sensitivity better than $10^{-10}$ emu. With this
sensibility we can obtain the magnetization loop for two NT of a
total mass of only $14$ pg. The temperature dependence and the
magnetization values indicate that the ferromagnetic alignment of
the moments are affected by the grain surface, where the spin
coupling is reduced from that at the core of each particle. The
results are consistent with those of a ferromagnetic material with
the shape anisotropy given by a cylindrical geometry. The measured
$H_{c}$ suggests the existence of magnetization reversal processes
that can overcome the energy barrier given by this anisotropy
constant.

\section{ACKNOWLEDGMENTS}

This work was partially supported by ANPCyT grant PICT04-03-21372.
M. I. D., D. A. and J. C. fellowship holders of CONICET. R. D. S.
and H. P. research members of CONICET. We thank F. de la Cruz for
a careful reading of the manuscript.


\clearpage

\begin{figure}
\begin{center}
\includegraphics[width=8.5cm]{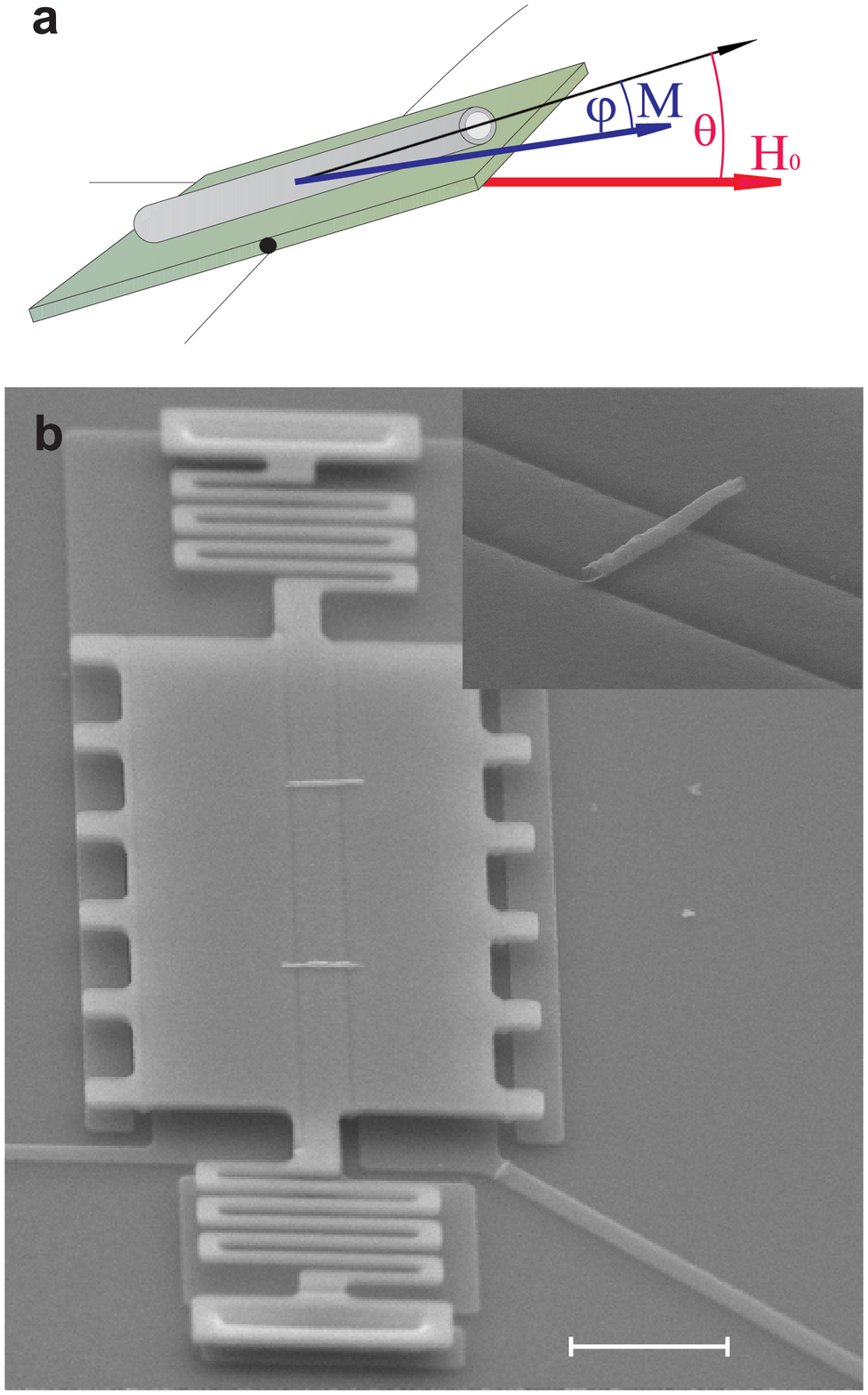}
\end{center}
\caption{ (a) Sketch of the different angles used in the analysis
of the measurements. (b) Scanning electron microscope image of the
LCMO nanotubes mounted on top of the polysilicon micro-oscillator.
The image was taken at an angle of 50 degrees to enhance the
topographic contrast. The dimension scale bar corresponds to
$20\,\mu$m. The inset shows a zoom in one of the nanotubes.}
\end{figure}

\begin{figure}
\begin{center}
\includegraphics[width=8.5cm,clip=true]{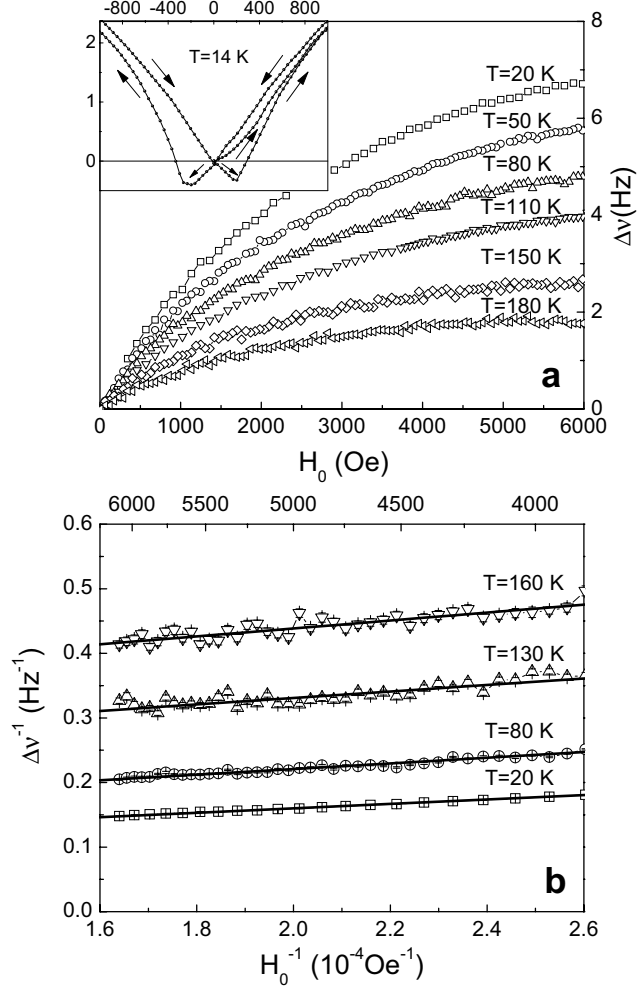}
\end{center}
\caption{ (a) Change in the resonant frequency as a function of
the external magnetic field for different temperatures as
indicated in the figure. The inset shows a zoomed area close to
zero magnetic field for a temperature of 14\,K. The arrows
indicate the direction of the magnetic field sweep in each branch
of the measurement. (b) Inverse of the frequency change as a
function of the inverse of the applied magnetic field for high
$H_0$ values is plotted for some selected temperatures (indicated
in the graph). The solid lines are fits of the data, which allows
to obtain the saturation magnetization and the shape anisotropy
constant. }
\end{figure}

\begin{figure}
\begin{center}
\includegraphics[width=8.5cm,clip=true]{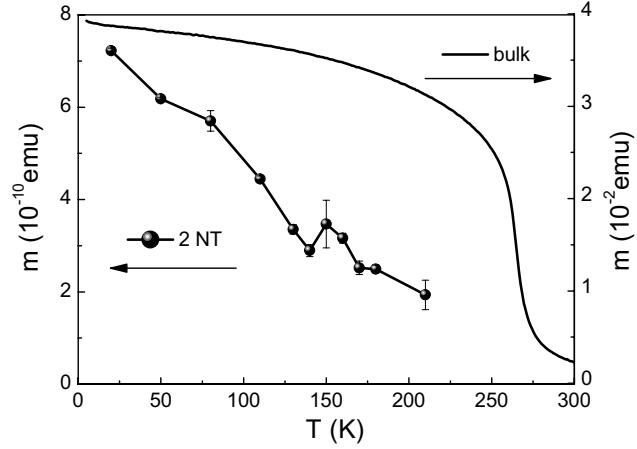}
\end{center}
\caption{ Saturation magnetization as a function of temperature.
Bullets are the data obtained from the measurements on two
isolated LCMO nanotubes. Solid line represents the magnetization
of LCMO bulk sample for an applied magnetic field of $10\,$kOe.}
\end{figure}

\begin{figure}
\begin{center}
\includegraphics[width=8.5cm,clip=true]{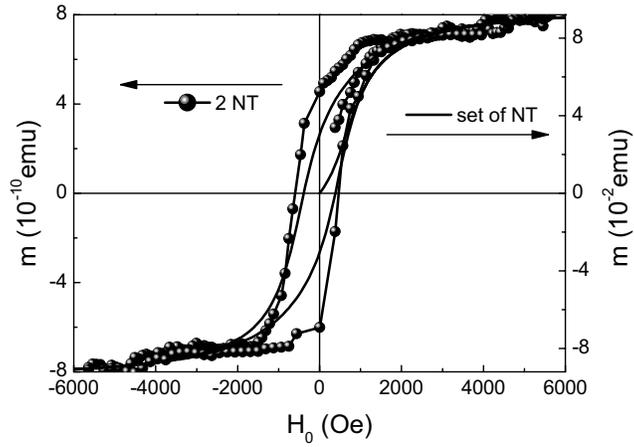}
\end{center} \caption{ Comparison between the
hysteresis loop of two isolated LCMO nanotubes obtained with the
torsional micro-oscillator magnetometer (bullets) and the loop
data of $1.84\,$mg of randomly oriented nanotubes taken with a
commercial SQUID magnetometer (continuous line).}
\end{figure}

\end{document}